\documentclass[a4paper,conference]{IEEEtran}
\usepackage{soul}
\usepackage[mathscr]{eucal}
\usepackage[cmex10]{amsmath}
\usepackage{epsfig,epsf,psfrag}
\usepackage{amssymb,amsmath,amsthm,amsfonts,latexsym}
\usepackage{amsmath,graphicx,bm,xcolor,url,overpic}
\usepackage[caption=false]{subfig} 
\usepackage{fixltx2e}
\usepackage{array}
\usepackage{verbatim}
\usepackage{bm}
\usepackage{algorithmic}
\usepackage{algorithm}
\usepackage{verbatim}
\usepackage{textcomp}
\usepackage{mathrsfs}
\usepackage{epstopdf}

\newcommand{\openone}{\leavevmode\hbox{\small1\normalsize\kern-.33em1}}

\catcode`~=11 \def\UrlSpecials{\do\~{\kern -.15em\lower .7ex\hbox{~}\kern .04em}} \catcode`~=13 

\allowdisplaybreaks[4]

\newcommand{\nn}{\nonumber}

\newcommand{\calA}{\mathcal{A}}

\newcommand{\calC}{\mathcal{C}}

\newcommand{\calE}{\mathcal{E}}
\newcommand{\calF}{\mathcal{F}}

\newcommand{\calP}{\mathcal{P}}

\newcommand{\calS}{\mathcal{S}}

\newcommand{\calX}{\mathcal{X}}
\newcommand{\calY}{\mathcal{Y}}

\newcommand{\bx}{\mathbf{x}}
\newcommand{\bX}{\mathbf{X}}

\newcommand{\rmA}{\mathrm{A}}

\newcommand{\rmc}{\mathrm{c}}

\newcommand{\rme}{\mathrm{e}}

\newcommand{\rmm}{\mathrm{m}}

\newcommand{\rmq}{\mathrm{q}}

\newcommand{\rmR}{\mathrm{R}}
\newcommand{\rms}{\mathrm{s}}

\newcommand{\bbE}{\mathbb{E}}

\newcommand{\bbN}{\mathbb{N}}

\newcommand{\bbP}{\mathbb{P}}

\newcommand{\bbR}{\mathbb{R}}

\DeclareMathAlphabet{\mathbsf}{OT1}{cmss}{bx}{n}
\DeclareMathAlphabet{\mathssf}{OT1}{cmss}{m}{sl}

\DeclareSymbolFont{bsfletters}{OT1}{cmss}{bx}{n}  
\DeclareSymbolFont{ssfletters}{OT1}{cmss}{m}{n}
\DeclareMathSymbol{\bsfGamma}{0}{bsfletters}{'000}
\DeclareMathSymbol{\ssfGamma}{0}{ssfletters}{'000}
\DeclareMathSymbol{\bsfDelta}{0}{bsfletters}{'001}
\DeclareMathSymbol{\ssfDelta}{0}{ssfletters}{'001}
\DeclareMathSymbol{\bsfTheta}{0}{bsfletters}{'002}
\DeclareMathSymbol{\ssfTheta}{0}{ssfletters}{'002}
\DeclareMathSymbol{\bsfLambda}{0}{bsfletters}{'003}
\DeclareMathSymbol{\ssfLambda}{0}{ssfletters}{'003}
\DeclareMathSymbol{\bsfXi}{0}{bsfletters}{'004}
\DeclareMathSymbol{\ssfXi}{0}{ssfletters}{'004}
\DeclareMathSymbol{\bsfPi}{0}{bsfletters}{'005}
\DeclareMathSymbol{\ssfPi}{0}{ssfletters}{'005}
\DeclareMathSymbol{\bsfSigma}{0}{bsfletters}{'006}
\DeclareMathSymbol{\ssfSigma}{0}{ssfletters}{'006}
\DeclareMathSymbol{\bsfUpsilon}{0}{bsfletters}{'007}
\DeclareMathSymbol{\ssfUpsilon}{0}{ssfletters}{'007}
\DeclareMathSymbol{\bsfPhi}{0}{bsfletters}{'010}
\DeclareMathSymbol{\ssfPhi}{0}{ssfletters}{'010}
\DeclareMathSymbol{\bsfPsi}{0}{bsfletters}{'011}
\DeclareMathSymbol{\ssfPsi}{0}{ssfletters}{'011}
\DeclareMathSymbol{\bsfOmega}{0}{bsfletters}{'012}
\DeclareMathSymbol{\ssfOmega}{0}{ssfletters}{'012}

\newcommand{\tilS}{\tilde{S}}

\DeclareMathOperator*{\argmax}{arg\,max}

\newtheorem{theorem}{Theorem}

\newtheorem{definition}{Definition}

\graphicspath{{./figures/}}
\usepackage{graphicx,cite}
\usepackage{epstopdf}
\usepackage{enumerate}
\usepackage{color}
\usepackage{bbm}
\usepackage{algorithm}
\usepackage{algorithmic}
\usepackage[T1]{fontenc}
\usepackage{aecompl}
\usepackage{booktabs}
\usepackage{tabularx}
\usepackage{bm}
\usepackage{dsfont}

\def\BibTeX{{\rm B\kern-.05em{\sc i\kern-.025em b}\kern-.08em T\kern-.1667em\lower.7ex\hbox{E}\kern-.125emX}}

\IEEEoverridecommandlockouts

\makeatletter

\newcommand{\Rmnum}[1]{\expandafter\@slowromancap\romannumeral #1@}
\makeatother

\def\BibTeX{{\rm B\kern-.05em{\sc i\kern-.025em b}\kern-.08em
T\kern-.1667em\lower.7ex\hbox{E}\kern-.125emX}}
\begin{document}
\title{Privacy-Resolution Tradeoff for Adaptive Noisy Twenty Questions Estimation}

\author{
\IEEEauthorblockN{Chunsong Sun and Lin Zhou} \\
\thanks{C. Sun is with the School of Cyber Science and Technology, Beihang University, Beijing, China, 100191, (Email: sunchunsong@buaa.edu.cn). L. Zhou is with School of Automation and Intelligent Manufacturing, Southern University of Science and Technology, Shenzhen, China, 518055, (Email: zhoul9@sustech.edu.cn).}
}

\maketitle

\begin{abstract}

We revisit noisy twenty questions estimation and study the privacy-resolution tradeoff for adaptive query procedures. Specifically, in twenty questions estimation, there are two players: an oracle and a questioner. The questioner aims to estimate target variables by posing queries to the oracle that knows the variables and using noisy responses to form reliable estimates. Typically, there are adaptive and non-adaptive query procedures. In adaptive querying, one designs the current query using previous queries and their noisy responses while in non-adaptive querying, all queries are posed simultaneously. Generally speaking, adaptive query procedures yield better performance. However, adaptive querying leads to privacy concerns, which were first studied by Tsitsiklis, Xu and Xu (COLT 2018) and by Xu, Xu and Yang (AISTATS 2021) for the noiseless case, where the oracle always provides correct answers to queries. In this paper, we generalize the above results to the more practical noisy case, by proposing a two-stage private query procedure, analyzing its non-asymptotic and second-order asymptotic achievable performance and discussing the impact of privacy concerns. Furthermore, when specialized to the noiseless case, our private query procedure achieves better performance than above-mentioned query procedures (COLT 2018, AISTATS 2021).

\end{abstract}


\section{Introduction} \label{sec_intro}

Inspired by once popular parlor games, R\'enyi~\cite{renyi1961problem} and Ulam~\cite{ulam1991adventures} pioneered the study of noisy twenty questions estimation for querying-and-answering-based variable estimation. The problem consists of two parties: a questioner who aims to estimate the unknown target random variable, and an oracle who knows the random variable. The questioner poses queries to the oracle and estimates the target random variable using potentially noisy responses from the oracle. Depending on how queries are generated, the query procedures of the questioner are categorized as adaptive~\cite{naghshvar2015extrinsic,kaspi2017searching,chiu2019noisy} or non-adaptive~\cite{jedynak2012twenty,zhou2021achievable,zhou2021resolution,zhou2022resolution,sun2023achievable,zhou2023resolution,sun2025moving,zhou2025twentyq} query procedures. In adaptive querying, one designs the current query using previous queries and their noisy responses while in non-adaptive querying, all queries are posed simultaneously. Generally speaking, adaptive query procedures could yield better performance, which is named benefit of adaptivity. Theoretically, the benefit of adaptivity was first proved by Kaspi, Shayevitz and Javidi~\cite{kaspi2017searching} asymptotically and by Zhou and Hero~\cite{zhou2021resolution,zhou2021achievable} non-asymptotically. The studies of twenty questions estimation with random noise were recently summarized by Zhou and Hero~\cite{zhou2025twentyq}.

Despite superior performance, adaptive query procedures suffer from privacy concerns. For ease of exposition, assume that the target random variable is an integer from $1$ to $16$ and consider the noiseless case where the oracle never lies. One typical adaptive query procedure is bisection search~\cite{knuth1998art}, where one halves the search region after each query. Without loss of generality, assume that the target variable is $3$. Using bisection search, the queries would be ``Is the number greater than $8$?'', ``Is it greater than $4$?'', ``Is it greater than $2$?'', ``Is it greater than $3$?''. This way, the questioner could estimate the target variable correctly with a noiseless oracle. However, any malicious third party that knows the queries could infer that the target random variable is either $3$ or $4$. This is called the privacy concern. Motivated by privacy concerns of adaptive querying, Tsitsiklis, Xu and Xu~\cite{tsitsiklis2018private} studied the theoretical benchmarks for adaptive twenty questions estimation with an eavesdropper so that the questioner can reliably estimate the target variable with high probability while the eavesdropper fails to estimate the target variable with high probability. The result was later refined by Xu, Xu and Yang~\cite{xu2021optimal}.

Although insightful, the results of~\cite{tsitsiklis2018private,xu2021optimal} were restricted to the noiseless case where the oracle never lies. However, in many practical applications of twenty questions such as beam alignment~\cite{chiu2019active,sun2026beam}, the responses are usually noisy. Towards a further step for practical applications, in this paper, we generalize the achievability part of~\cite{tsitsiklis2018private,xu2021optimal} to the noisy case and reveal the privacy-resolution tradeoff. Specifically, we propose a two-stage private query procedure, which consists of a non-adaptive query stage and a parallel adaptive query stage. Subsequently, we analyze the non-asymptotic and second-order asymptotic performance of our private query procedure, inspired by the analysis of variable-length feedback codes~\cite{yavas2025variable} and the analysis of sorted posterior matching (sortPM)~\cite{chiu2021low}. Our results illustrate the impact of privacy concerns due to the eavesdropper on the reliability performance of the questioner. In particular, our results bridge over theoretical benchmarks between adaptive and non-adaptive query procedures. Furthermore, when specialized to the noiseless case, the performance of our private query procedure is better than the query procedures used in \cite{tsitsiklis2018private,xu2021optimal}.

\section{Problem Formulation} \label{sec_pf}

\subsection*{Notation}

Random variables and their realizations are denoted by upper case variables (e.g., $X$) and lower case variables (e.g., $x$), respectively. All sets are denoted in calligraphic font (e.g., $\calX$). We use $\bbR, \bbR_+$ and $\bbN$ to denote the set of real numbers, positive real numbers and integers, respectively. Given any real number $a\in\bbR$, let $a^+:=\max\{0,a\}$. For any random variable $X$ with distribution $P_X$ and $\bbE[X]>0$, the essential supremum of $X$ is defined as $\text{esssup}(X):=\sup\{a \in \bbR:\text{Pr}[X \geq a]>0\}$ and define a constant
\begin{align}
b(P_X):=\min\bigg\{\frac{\bbE[(X^+)^2]}{\bbE[X]},\text{esssup}(X)\bigg\} . \label{b-func}
\end{align}
Given any positive integer $n \in \bbN$, let $X^n:=(X_1,\ldots,X_n)$ be a random vector of length $n$. Given any two positive integers $(m,n) \in \bbN^2$, we use $[m:n]$ to denote the set $\{m,m+1,\ldots,n\}$, and $[m]$ to denote $[1 : m]$. All logarithms are base $e$. We use $\mathds{1}(\cdot)$ to denote the indicator function. Given two alphabets $(\calX,\calY)$, we use $\calF(\calX)$ to denote the set of all probability distributions on the set $\calX$ and use $\calP(\calY|\calX)$ to denote the set of all conditional probability distributions from $\calX$ to $\calY$. Given any $p \in [0,1]$, let Bern$(p)$ denote the Bernoulli distribution with parameter $p$. Given two distributions $(P,Q)\in\calF(\calX)^2$, we denote the Kullback-Leibler (KL) divergence by
\begin{align}
D(P\|Q) = \sum_{x \in \calX} P(x) \log \frac{P(x)}{Q(x)}. \label{KL-div}
\end{align}

\subsection{Adaptive Noisy Twenty Questions Estimation with Eavesdropper under Query-Dependent Channel} \label{pf System Model}

\begin{figure}[tb]
\centering
\includegraphics[width=.75\columnwidth]{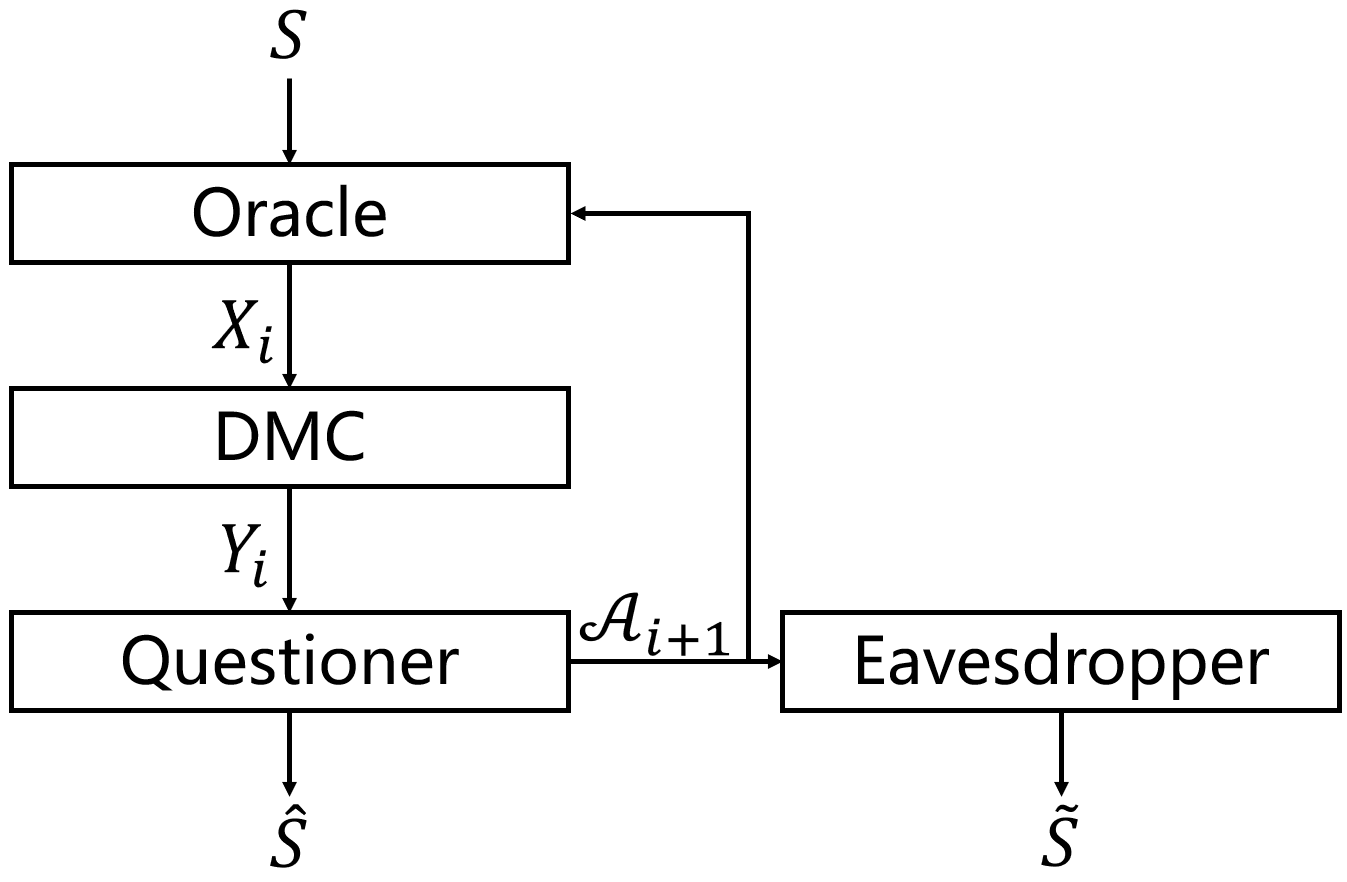}
\caption{System model for adaptive noisy twenty questions estimation with an eavesdropper.}
\label{adaptive-framework}
\end{figure}

As shown in Fig. \ref{adaptive-framework}, the questioner uses an adaptive query procedure to estimate a target random variable $S$ and an eavesdropper aims to infer the target random variable from the queries posed by the questioner. Formally, let $S\in[0,1]$ be a random variable generated from an arbitrary probability density function (pdf) $f_S$. At each time point $i \in \bbN$, the questioner generates a Lebesgue measurable query $\calA_i \subseteq [0,1]$ and queries the oracle whether $S$ lies in the set $\calA_i$. The query set $\calA_i$ is generated using all previous queries $\{\calA_1,\ldots,\calA_{i-1}\}$ and their noisy responses $(Y_1,\ldots,Y_{i-1})$ and is revealed to the eavesdropper. After receiving $\calA_i$, the oracle generates a binary yes/no answer $X_i=\mathds{1}(S\in\calA_i)$ and passes $X_i$ over a query-dependent discrete memoryless channel (DMC)~\cite[Eq. (1)]{zhou2021resolution} with transition probability $P_{Y_i|X_i}^{h(|\calA_i|)} \in \calP(\calY|\{0,1\})$, yielding a noisy response $Y_i$, where $h(\cdot)$ is a bounded Lipschitz continuous function, and $|\calA_i|$ denotes the size of the query $\calA_i$. Here the DMC models the behavior of the oracle, who can provide incorrect answers or decline to answer a query. Subsequently, using responses to all queries $(Y_1,\ldots,Y_i)$, the questioner determines whether to stop the query procedure. If yes, the questioner generates an estimate $\hat{S}$. Otherwise, the query procedure continues.

\subsection{Definition of Private Query Procedure} \label{pf Private Query Procedure}

To formally define the privacy-resolution tradeoff, we define a private query procedure as follows.

\begin{definition} \label{def_queryprocedure}
Given any pair of number of queries and privacy level $(N,L) \in \bbN^2$, any resolution of $\delta \in \bbR_+$, and any excess-resolution probability $\varepsilon \in [0, 1]$, an $(N,L,\delta,\varepsilon)$ private query procedure consists of
\begin{itemize}
\item[$\bullet$] a sequence of queries $\calA_i \subseteq [0,1]$ for each $i \in \bbN$,
\item[$\bullet$] a sequence of decoding function for the questioner $g_{\rmq,i}:\calY^i \rightarrow \calS \subseteq [0,1]$,
\item[$\bullet$] a sequence of decoding function for the eavesdropper $g_{\rme,i}:\calA^i \rightarrow \calS \subseteq [0,1]$,
\item[$\bullet$] a random stopping time $\tau$ depending on noisy responses $Y^i$ such that under any pdf $f_S$ of the random variable $S$, the average number of queries satisfies $\mathbb{E}[\tau] \leq N$,
\end{itemize}
which ensures that the excess-resolution probability for the questioner satisfies
\begin{align}
&\bbP_\rmq(N,\delta) := \sup_{f_S \in \calF(\calS)} \mathrm{Pr}\Big\{ \big| \hat{S}-S \big| > \delta \Big\} \leq \varepsilon, \label{p-q}
\end{align}
and for any $k \in [1,\lceil\frac{L}{2}\rceil]$, the probability of correct estimation of the eavesdropper satisfies
\begin{align}
&\bbP_\rme(N,L) := \sup_{f_S \in \calF(\calS)} \mathrm{Pr}\Big\{ \big| \tilS-S \big| \leq \frac{2k-1}{2L} \Big\} \leq \frac{2k-1}{L}, \label{p-e}
\end{align}
where $\hat{S}$ is the questioner's estimate of the target $S$ using the decoder $g_{\rmq,\tau}$, i.e., $g_{\rmq,\tau}(Y^{\tau})=\hat{S}$, while $\tilS$ is the eavesdropper's estimate using the decoder $g_{\rme,\tau}$, i.e., $g_{\rme,\tau}(\calA^{\tau})=\tilS$. 
\end{definition}

We remark that the parameter $L$ is introduced to quantify the privacy level of the eavesdropper and the constraint in \eqref{p-e} ensures that the eavesdropper could not estimate the target variable within resolution $\frac{2k-1}{2L}$ with probability higher than $\frac{2k-1}{L}$. The term $\frac{2k-1}{L}$ in \eqref{p-e} arises because i) when we evaluate the privacy concerns, we partition the unit interval $[0,1]$ into $L$ equal-sized non-overlapping sub-intervals; and ii) once the eavesdropper's estimate $\tilS$ lies in the same sub-interval as $S$ or in one of the $2k$ adjacent sub-intervals, the estimation could be within resolution $\frac{2k-1}{2L}$. When $L=1$, the constraint in \eqref{p-e} is always satisfied, which reduces to adaptive querying without the privacy constraint; when $L=\frac{1}{\delta}$\footnote{Without loss of generality, we relax the integer constraint on $\frac{1}{\delta}$.}, the above definitions reduce to non-adaptive querying. This is because when $L=\frac{1}{\delta}$, choosing $k=1$ leads to the constraint that 
\begin{align}
\bbP_\rme(N,L) := \sup_{f_S \in \calF(\calS)} \mathrm{Pr}\big\{ \big| \tilS-S \big| \leq 0.5\delta \big\} \leq \delta.
\end{align}
The above constraint could be satisfied by an eavesdropper who estimates the target random variable randomly and uniformly over $L$ sub-intervals. It implies that no information about the target is leaked via queries, which can only be achieved via non-adaptive querying. Therefore, to avoid degenerate cases, we focus on the regime $L\in [2,\lfloor\frac{1}{\delta}\rfloor]$.

The theoretical benchmark of interest is defined as the following minimal achievable resolution subject to a privacy constraint:
\begin{align}
\delta^*(N,L,\varepsilon)
\nn:=\inf\{\delta:\exists~\mathrm{an}~(n,L,&\delta,\varepsilon)\mathrm{-private}\\*
&\mathrm{~query~procedure}\}. \label{delta_star}
\end{align}

\section{Main Results} \label{sec_mr}

\subsection{Preliminaries} \label{mr Preliminary Definitions}

This subsection presents definitions needed to present our query procedure and theoretical results. Fix any integer $M\in\bbN$. Assume that the unit interval $[0,1]$ is divided into $M$ equal-sized non-overlapping sub-intervals $(\calC_1,\ldots,\calC_M)$.

Our private query procedure consists of two stages. The first stage is non-adaptive and the second stage is adaptive. In the following, we describe the first stage, the second stage and the overall query procedure, respectively.

\subsubsection{Non-adaptive query procedure}
In the first stage, we use the non-adaptive query procedure in~\cite{zhou2021achievable} that is based on variable-length channel coding~\cite{polyanskiy2011feedback}. The key idea is to generate queries non-adaptively while allowing the number of queries to vary. In the following, we explain how the number of queries, a.k.a. the stopping time, is calculated since the query generation process is exactly the same as~\cite{zhou2021resolution,zhou2022resolution,zhou2023resolution}.

To do so, given any $p \in [0,1]$, define the mutual information density as follows:
\begin{align}
\imath^{h(p)}(x;y):=\log\frac{P_{Y|X}^{h(p)}(y|x)}{P_{Y}^{h(p)}(y)},~\forall (x,y) \in \calX \times \calY, \label{mutual-info-density}
\end{align}
where $P_{Y|X}^{h(p)}$ denotes the query-dependent channel and $P_{Y}^{h(p)}$ denotes the marginal distribution on $\calY$ induced by $P_{Y|X}^{h(p)}$ and the Bernoulli distribution $P_X=\text{Bern}(p)$. Given any integer $n \in \bbN$, for any $(x^n,y^n) \in \calX^n \times \calY^n$, define the mutual information density of two sequences $x^n$ and $y^n$ as follows:
\begin{align}
\imath^{h(p)}(x^n;y^n):=\sum_{i \in [n]} \imath^{h(p)}(x_i;y_i).
\end{align}
Let $\bX^{\infty}=(X^\infty(1),\ldots,X^\infty(M))$ be a sequence of $M$ binary codewords with infinite length and $Y^{\infty}$ be the noisy responses. Let $\bx^{\infty}$ denote a realization of $\bX^{\infty}$ and $y^{\infty}$ denote a realization of $Y^{\infty}$. Given any threshold $\lambda \in \bbR_+$ and any $j \in [M]$, define the following stopping time: 
\begin{align}
\tau_{\rmm,j}(\bx^{\infty},y^{\infty},\lambda):=\inf\big\{ n \in \bbN:\imath^{h(p)}(x^n(j);y^n) \geq \lambda \big\}. \label{tau-m}
\end{align}
Finally, the minimal stopping time is defined as follows:
\begin{align}
\tau_{\rmm}^*(\bx^{\infty},y^{\infty},\lambda) := \min_{j \in [M]} \tau_{\rmm,j}(\bx^{\infty},y^{\infty},\lambda). \label{tau-m-star}
\end{align}
For ease of notation,  let $\tau_{\rmm}^*$ denote $\tau_{\rmm}^*(\bx^{\infty},y^{\infty},\lambda)$. After $\tau_{\rmm}^*$ queries, we stop and obtain an estimate that the target lies within the $j_{\rmm}$-th sub-intervals, where
\begin{align}
j_{\rmm} = \max\big\{j \in [M]:\imath^{h(p)}\big(x^{\tau_{\rmm}^*}(j);y^{\tau_{\rmm}^*}\big) \geq \lambda\big\}.
\end{align}

\subsubsection{Adaptive query procedure}
In the second stage, we use the adaptive query procedure in~\cite{chiu2021low}. Let $\bm{\rho}(i)=(\rho_1(i),\ldots,\rho_{M}(i))$ denote the vector of posterior probabilities, which quantifies the probability that the target random variable lies in each sub-interval at each time point $i \in \bbN$. Initialize $\bm{\rho}(0)$ as a uniform Bayesian prior, i.e., $\rho_j(0)=\frac{1}{M}$ for all $j\in[M]$. Since the query generation process is exactly the same as~\cite{chiu2021low}, here we only explain how the posterior probabilities are updated and how the stopping time is calculated.

At each time point $i \in \bbN$, after posing the query set $\calA_i$, the questioner receives a noisy response $Y_i$. Using the query set $\calA_i$ and the noisy response $Y_i$, for any $j \in [M]$, the posterior probability $\rho_j(i)$ can be updated as follows:
\begin{align}
\rho_j(i) =\frac{\rho_j(i-1)g(Y_i|\mathds{1}\{s \in \calA_i\},\mathds{1}\{\calC_j \subseteq \calA_i\})}{\sum_{j^{\prime}\in[M]}\rho_{j^{\prime}}(i-1)g(Y_i|\mathds{1}\{s \in \calA_i\},\mathds{1}\{\calC_{j^{\prime}} \subseteq \calA_i\})}. \label{update-prob}
\end{align}
where $g(Y_i|\mathds{1}\{s \in \calA_i\},\mathds{1}\{\calC_j \subseteq \calA_i\})$ is the conditional distribution of $Y_i$ given the noiseless response $\mathds{1}\{s \in \calA_i\}$ and whether the query $\calA_i$ contains the sub-interval $\calC_j$.

Given any $\varepsilon^{\prime} \in [0,1]$, for any $j \in [M]$, define the following stopping time:
\begin{align}
\tau_{\rms,j}(\bx^{\infty},y^{\infty},\varepsilon^{\prime}):=\inf\big\{ n \in \bbN: \rho_j(n) \geq 1-\varepsilon^{\prime} \big\}, \label{tau-s}
\end{align}
and define the minimal stopping time as follows:
\begin{align}
\tau_{\rms}^*(\bx^{\infty},y^{\infty},\varepsilon^{\prime}):=\min_{j \in [M]} \tau_{\rms,j}(\bx^{\infty},y^\infty,\varepsilon^{\prime}). \label{tau-s-star}
\end{align}
For ease of notation, let $\tau_{\rms}^*$ denote $\tau_{\rms}^*(\bx^{\infty},y^{\infty},\varepsilon^{\prime})$. After $\tau_{\rms}^*$ queries, we stop and obtain an estimate that the target lies within the $j_{\rms}$-th sub-intervals, where
\begin{align}
j_{\rms} = \argmax_{j \in [M]} \rho_j(\tau_{\rms}^*).
\end{align}

\subsection{Our Two-Stage Private Query Procedure} \label{mr Three-Stage Private Query Procedure}

\begin{figure}[tb]
\centering
\includegraphics[width=1.\columnwidth]{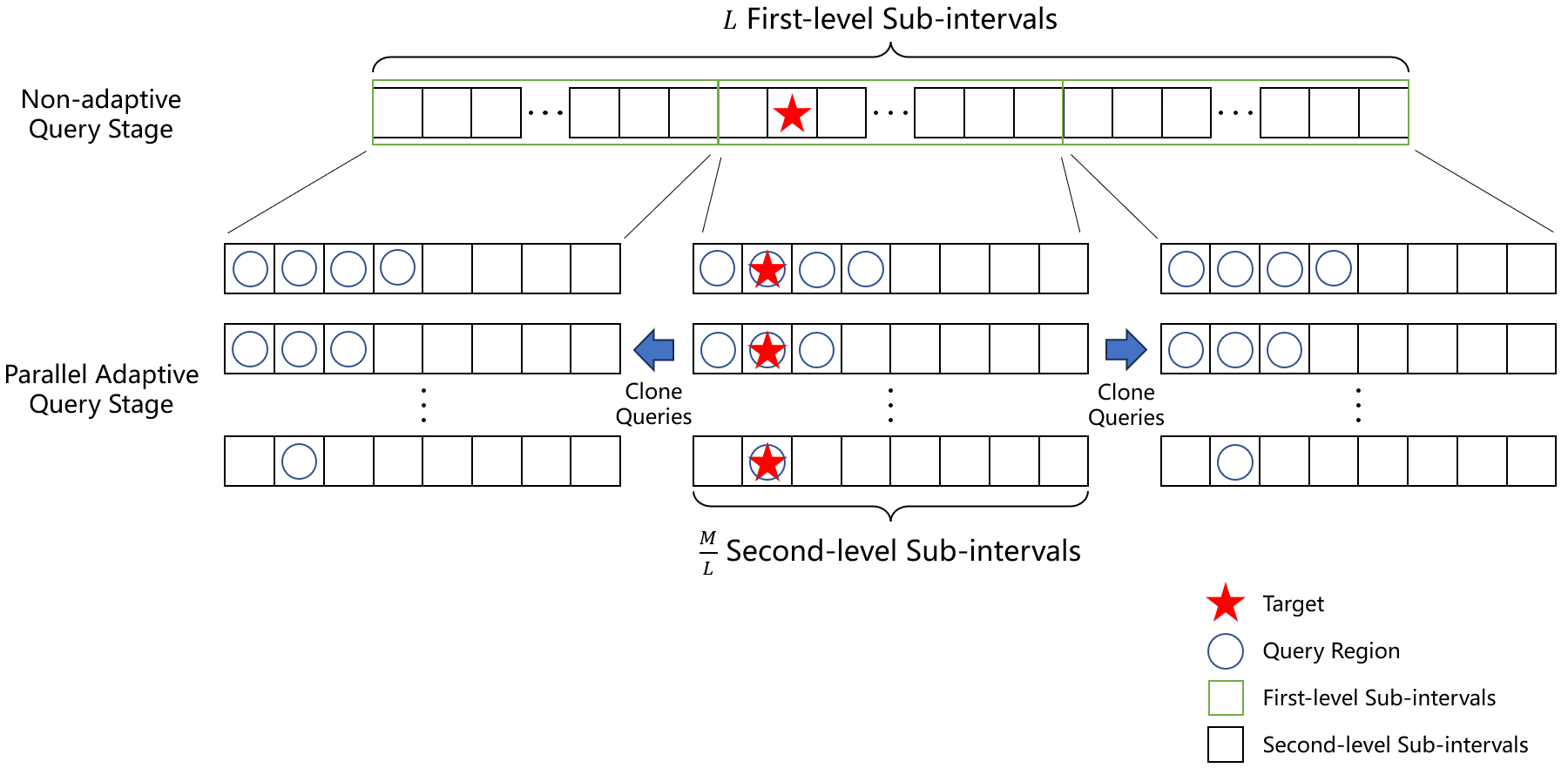}
\caption{Illustration of our two-stage private query procedure.}
\label{twoStage}
\end{figure}

To tradeoff privacy against the eavesdropper and the achievable resolution of the questioner, as shown in Fig. \ref{twoStage}, we propose a two-stage private query procedure that consists of a non-adaptive query stage and a parallel adaptive query stage. Fix positive integers $(N_0,M) \in \bbN^2$, a positive integer $L \in [2,M-1]$, and positive real numbers $(\varepsilon^{\prime},\lambda_1,\lambda_2)\in[0,1]\times\bbR_+^2$ satisfying $\lambda_1<\lambda_2$. Let $(N_1,N_2) \in \bbN^2$ denote the number of queries in the non-adaptive query stage and the parallel adaptive query stage, respectively. In the non-adaptive query stage, the questioner uses at most two variable-length query procedures, which are connected via a hypothesis test.

In the following, we describe how our private adaptive query procedure works. Partition the unit interval $[0,1]$ into $L$ equally sized non-overlapping first-level sub-intervals $(\calC_1^{(1)},\ldots,\calC_L^{(1)})$. The questioner uses the random coding strategy to generate $L$ binary codewords  $(x^\infty(1),\ldots,x^\infty(L))$ of infinite length,  which are used to form queries as~\cite[Algorithm 1]{zhou2021resolution}. Let $y^{\infty}$ be the noisy responses. After $i \in \bbN$ queries, if
\begin{align}
\max_{j\in[L]} \imath^{h(p)}\big(x^{i}(j);y^{i}\big)\geq\lambda_1,
\end{align}
then the questioner stops and obtains a first estimate of the first-level sub-interval in which the target $S$ lies, denoted by $W_1^{(1)}$. Recall the definition of $D(\cdot)$ in \eqref{KL-div} in Notation part. Let
\begin{align}
(x_{\rmA},x_{\rmR}):=\argmax_{(x,x^{\prime})\in\calX^2} D\big(P_{Y|X=x}^{h(p)}\big\|P_{Y|X=x^{\prime}}^{h(p)}\big).
\end{align}
Similar to~\cite[Section VI-A]{yavas2025variable}, the questioner conducts a sequential hypothesis test, with hypothesis $H_{\rmA}$ corresponding to accepting $W_1^{(1)}$ and hypothesis $H_{\rmR}$ corresponding to rejection. If $H_{\rmA}$ is declared, the estimate in this stage is set as $W^{(1)}$. Otherwise, if $H_{\rmR}$ is declared, the questioner continues to pose queries starting from index $(i+1)$, until the mutual information density of a first-level sub-interval exceeds a threshold $\lambda_2$. Subsequently, the questioner obtains a second estimate $W_2^{(1)}$ and sets $W^{(1)}$. In either case, the number of queries is denoted as $N_1$.

In the second stage, the first-level sub-interval $\calC_{W^{(1)}}^{(1)}$ is further divided into $\lceil \frac{M}{L} \rceil$ equal-sized non-overlapping second-level sub-intervals $(\calC_1^{(2)},\ldots,\calC_{\lceil M/L \rceil}^{(2)})$. Without loss of generality, let $\lceil \frac{M}{L} \rceil=\frac{M}{L}$. The questioner forms the query $\calA_i$ using sortPM strategy~\cite[Algorithm 1]{chiu2021low}, and updates the posterior probabilities by \eqref{update-prob} with noisy response $Y_i$. Furthermore, to confuse the eavesdropper, the questioner clones all the queries in parallel on each $L-1$ first-level sub-intervals, as shown in Fig. \ref{twoStage}. Since the questioner has already known $W^{(1)}$, this cloning operation does not affect the estimation in this stage. But the cloning operation ensures that the eavesdropper cannot obtain any information about $W^{(1)}$. After $N_2$ queries, if
\begin{align}
\max_{j\in[M/L]}\rho_j(N_2)>1-\varepsilon^{\prime},
\end{align}
the questioner stops and obtains a estimate of the second-level sub-interval in which the target $S$ lies, denoted by $W^{(2)}$. Finally, the questioner takes the center of $\calC_{W^{(2)}}^{(2)}$ as the estimate $\hat{S}$, and the total number of required queries is $N_1+N_2$.

\subsection{Results, Discussions and Numerical Illustrations}

In the following theorem, we derive non-asymptotic bounds for the two-stage private query procedure. Let $\bX_2^{\infty}$ be a sequence of $\frac{M}{L}$ binary codewords generated in the parallel adaptive query stage, and let $Y_2^{\infty}$ be the corresponding noisy responses. Recall the definition of $\imath^{h(p)}(x;y)$ in \eqref{mutual-info-density}. Define the ``capacity'' of the binary-input query-dependent channel $P_{Y|X}^{h(p)}$ as follows:
\begin{align}
C=\max_{p \in [0,1]}\bbE\big[\imath^{h(p)}(X;Y)\big]. \label{capacity}
\end{align}

\begin{theorem} \label{nonasymptotic}
Given any $M \in \bbN$ and any $L \in [2,M-1]$, for any $(p,\varepsilon_0,\varepsilon^{\prime},\lambda_1,\lambda_2) \in [0,1]^3 \times \bbR_+^2$, there exists an $(N,L,1/M,\varepsilon)$-private query procedure such that
\begin{align}
N &\leq (1-\varepsilon_0)\bar{N}, \\
\varepsilon &\leq \varepsilon_0+(1-\varepsilon_0)\bar{\varepsilon}, \label{nonasymptotic-rlt}
\end{align}
where
\begin{align}
&\bar{N}=\frac{\lambda_1+b}{C}+\big((L-1)\exp(-\lambda_1) + \exp(-a_{\rmR})\big)\frac{\lambda_2-\lambda_1+b}{C} \nn\\
&+ \frac{a_{\rmA}+b_{\rmA}}{D\big(P_{Y|X=x_{\rmA}}^{h(p)}\big\|P_{Y|X=x_{\rmR}}^{h(p)}\big)} \nn\\
&+ (L-1)\exp(-\lambda_1)\frac{a_{\rmR}+b_{\rmR}}{D\big(P_{Y|X=x_{\rmR}}^{h(p)}\big\|P_{Y|X=x_{\rmA}}^{h(p)}\big)}, \nn \\
&+\bbE\big[\tau_{\rms,1}(\bX_2^{\infty},Y_2^{\infty},\varepsilon^{\prime})\big] \nn\\
&+ (L-1)\big(\exp(-\lambda_1+a_{\rmA})+\exp(-\lambda_2)\big)N_0, \label{nonasy-N}\\
&\bar{\varepsilon} = (L-1)\big(\exp(-\lambda_1-a_{\rmA})+\exp(-\lambda_2)\big)+\varepsilon^{\prime}. \label{nonasy-var}
\end{align}
\end{theorem}

To prove Theorem \ref{nonasymptotic}, we combine the proposed two-stage private query stage with the stop-at-time-zero strategy~\cite{polyanskiy2011feedback}, with the stopping probability $\varepsilon_0$. We next briefly explain what each term in Theorem \ref{nonasymptotic} means. In \eqref{nonasy-N}, the first four terms bound the average stopping time in the non-adaptive query stage, and the last two terms bound the average stopping time in the parallel adaptive query stage. Similarly, in \eqref{nonasy-var}, the first and second terms correspond to the error probabilities in the non-adaptive query stage and the parallel adaptive query stage, respectively.

Recall that $P_{Y|X}^{h(p)}$ denotes the query-dependent noisy channel. Let
\begin{align}
\widetilde{C} = \max_{(x,x^{\prime}) \in \calX^2} D\big(P_{Y|X=x}^{h(p)}\big\|P_{Y|X=x^{\prime}}^{h(p)}\big) \label{capacity-wt}
\end{align}
represent the maximal KL divergence between the conditional output distributions of any two input symbols. We have the following second-order asymptotic result.

\begin{theorem} \label{2order-asymptotic}
For any $(N,L,\varepsilon) \in \bbN^2\times [0,1]$ such that $L\geq 2$, the minimal achievable resolution $\delta^*(N,L,\varepsilon)$ of an optimal private query procedure satisfies
\begin{align}
&-\log \delta^*(N,L,\varepsilon) \nn\\*
&\geq \frac{1}{1-\varepsilon}CN-\frac{C}{\widetilde{C}}\log N_1-\log\log N_1-\frac{C}{\widetilde{C}}\log\frac{1}{\varepsilon^{\prime}}+O(1), \label{2order-delta}
\end{align}
where $N_1$ is the solution to the following equality
\begin{align}
\log L = CN_1-\frac{C}{\widetilde{C}}\log N_1-\log\log N_1. \label{logL-N1}
\end{align}
\end{theorem}
The proof of Theorem \ref{2order-asymptotic} generalizes the proofs in~\cite[Theorem 2]{yavas2025variable} and~\cite[Theorem 1]{chiu2021low} and is available in our extended version~\cite{sun2026privacy}.

\begin{figure}[tb]
\centering
\includegraphics[width=.75\columnwidth]{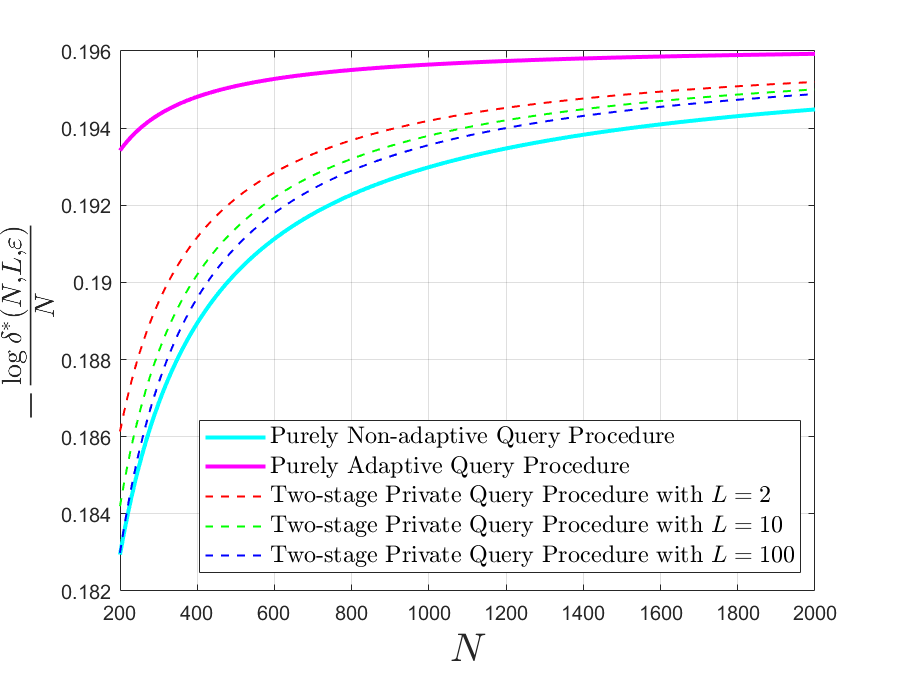}
\caption{Plot of achievability bounds on the resolution decay rate $-\frac{\log \delta^*(N,L,\varepsilon)}{N}$ as a function of the number of queries $N$ for various privacy levels $L$ with the excess-resolution probability $\varepsilon=0.1$, under a query-dependent binary symmetric channel with $h(p)=0.1+0.3p$.}
\label{adaptivityGain-vs-security}
\end{figure}

To demonstrate the impact of privacy concerns due to the eavesdropper on the reliability performance of the questioner, in Fig. \ref{adaptivityGain-vs-security}, we numerically compare the asymptotic results of the proposed two-stage private query procedure (dashed line in orange, red, green and blue), the purely non-adaptive query procedure~\cite{yavas2025variable} (solid line in cyan) and the purely adaptive query procedure~\cite{chiu2021low} (solid line in magenta). As observed, our private adaptive query procedure bridges over adaptive and non-adaptive query procedures via the privacy level $L$.

\begin{figure}[tb]
\centering
\includegraphics[width=.75\columnwidth]{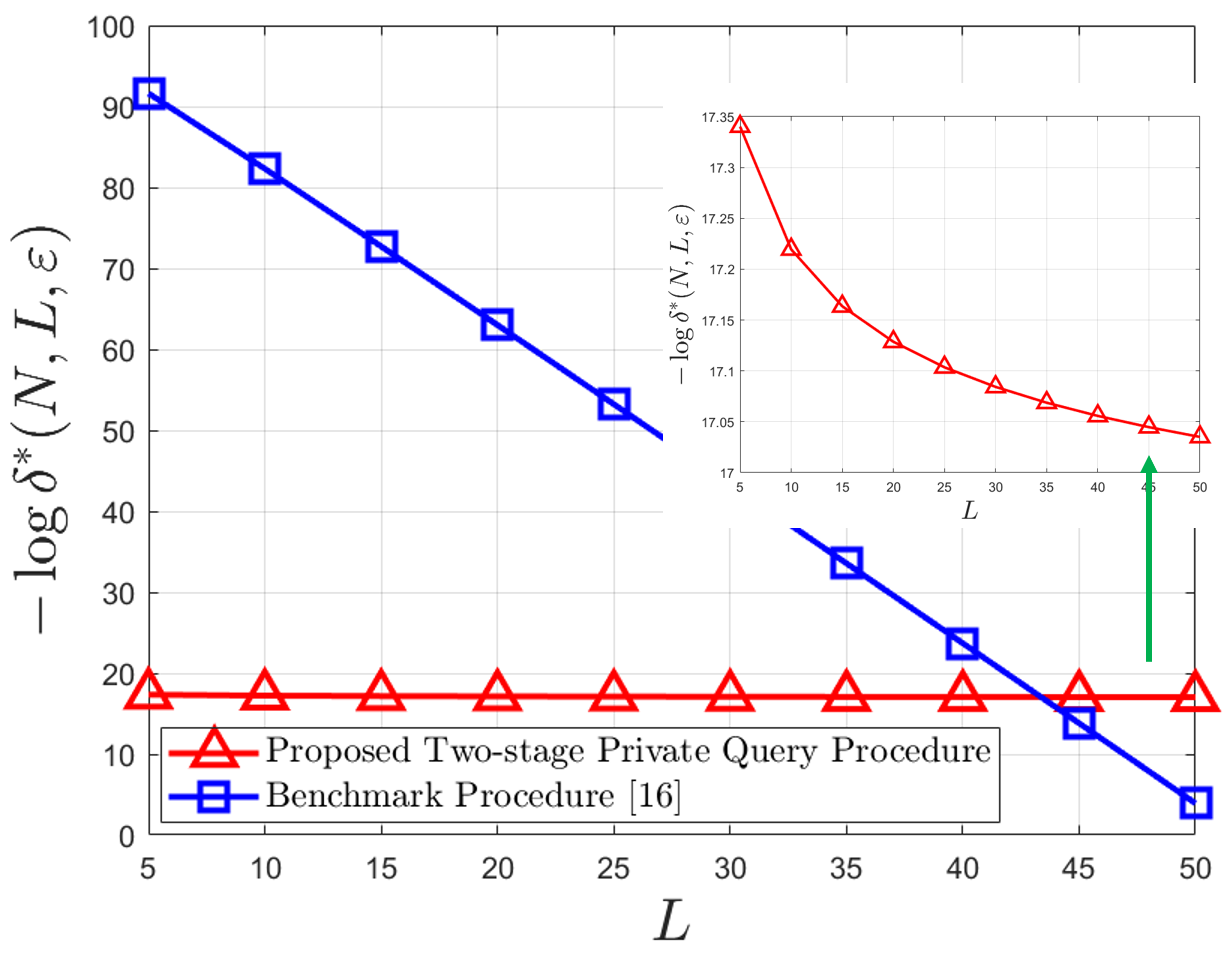}
\caption{Plot of the achievable resolutions $-\log \delta^*(N,L,\varepsilon)$ of the proposed two-stage private query procedure and the benchmark procedure~\cite{xu2021optimal}, as a function of the privacy level $L$ for the fixed number of queries $N=100$. For the noisy case, the excess-resolution probability is set to $\varepsilon=0.1$, and the channel is a query-dependent binary symmetric channel with $h(p)=0.1+0.3p$.}
\label{noiseless-vs-noisy}
\end{figure}

\begin{figure}[tb]
\centering
\includegraphics[width=.75\columnwidth]{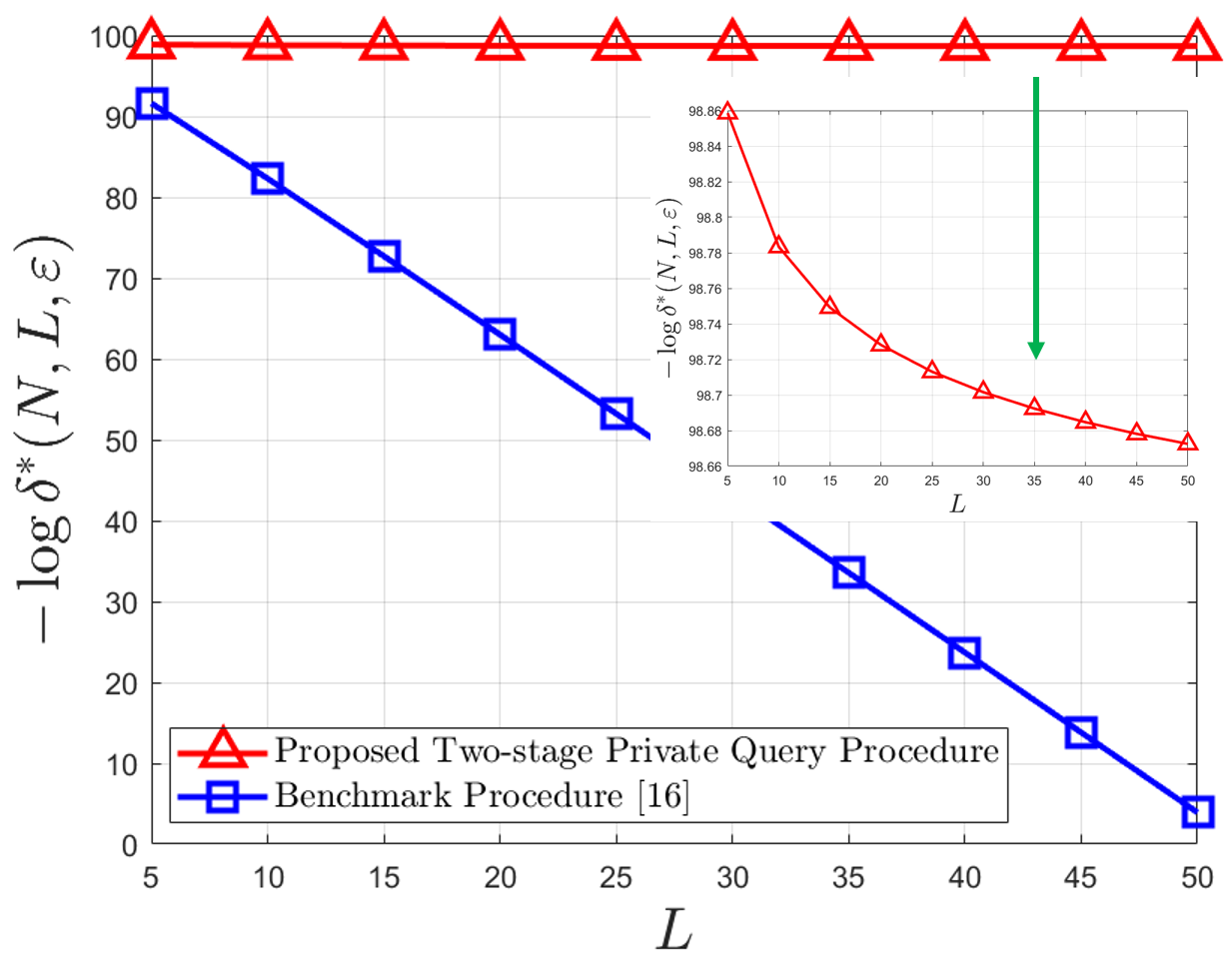}
\caption{Plot of the achievable resolutions $-\log \delta^*(N,L,\varepsilon)$ of the proposed two-stage private query procedure and the benchmark procedure~\cite{xu2021optimal}, as a function of the privacy level $L$ for the fixed number of queries $N=100$, when specialized to the noiseless case.}
\label{both-noiseless}
\end{figure}

We next compare our results for the noisy case with the benchmark in \cite{xu2021optimal} for the noiseless case to demonstrate the impact of noise and to show the performance gap. It follows from \cite[Theorem 2]{xu2021optimal} that $-\delta_{\text{nl}}^*(N,L)=N+\log L-2L+O(1)$. To demonstrate the gap between the noisy and noiseless cases, in Fig. \ref{noiseless-vs-noisy}, we numerically compare the achievable resolution $-\log \delta^*(N,L,\varepsilon)$ as a function of the privacy level $L$ for the fixed number of queries $N=100$. As observed, the benchmark procedure in \cite[Section 4.2]{xu2021optimal} is highly sensitive to variations in the privacy level $L$, while our query procedure maintains relatively stable. This is because for our query procedure, when $N$ is fixed, the achievable resolution is only affected by the second and third terms on the right-hand side of \eqref{2order-delta}. It follows from \eqref{logL-N1} that $\log N_1=O(\log\log L)$, which implies the achievable resolution does not vary significantly with respect to the privacy level $L$. Finally, when specializing our query procedure to the noiseless case, we numerically verify in Fig. \ref{both-noiseless} that our query procedure achieves better performance than the benchmark procedure in \cite[Section 4.2]{xu2021optimal}.

\section{Conclusion}

We revisited adaptive twenty questions estimation, proposed a two-stage private query procedure and derived non-asymptotic and second-order asymptotic achievability bounds. Furthermore, we numerically illustrated the impact of the privacy level and the noise level. Finally, when specialized to the noiseless case, we showed that our query procedure achieved better performance than the benchmark procedure in~\cite[Section 4.2]{xu2021optimal}. In the future, one could derive converse bounds or explore applications including active beam alignment~\cite{chiu2019active,sohrabi2021deep,sun2026beam} and best arm identification~\cite{Cohen2025MAB}.

\clearpage

\bibliographystyle{IEEEtran}
\bibliography{IEEEabrv,IEEE_cs}

\clearpage

\appendix

\subsection{Proof of the Non-Asymptotic Bound (Theorem \ref{nonasymptotic})} \label{appendix_a}

Fix positive integers $(N_0,M) \in \bbN^2$, positive integer $L \in [2,M-1]$, positive real numbers $(\varepsilon^{\prime},\lambda_1,\lambda_2)\in[0,1]\times\bbR_+^2$ satisfying $\lambda_1<\lambda_2$. Let $\bX_1^{\infty}:=(X_1^{\infty}(1),\ldots,X_1^{\infty}(L))$ be a sequence of binary codewords with length-$L$ in the non-adaptive query stage, where each codeword is generated i.i.d from the Bernoulli distribution $P_X$ with parameter $p \in [0,1]$. Let $\bX_2^{\infty}:=(X_2^{\infty}(1),\ldots,X_2^{\infty}(\frac{M}{L}))$ be a sequence of binary codewords with length-$\frac{M}{L}$ in the cloned adaptive query stage, where each codeword is generated by the sortPM strategy mentioned in Section III-A. Furthermore, let $Y_1^{\infty}$ and $Y_2^{\infty}$ be the random vector with infinite length for the first and second stages, respectively, where each element takes values in $\calY$. Fix $p \in [0,1]$. Unless otherwise specified, we use $P_{Y|X}$ to denote $P_{Y|X}^{h(p)}$ in what follows.

\textbf{The first estimation in the non-adaptive query stage:}
Without loss of generality, assume that the target $S \in [0,1]$ lies in the first- and second-level sub-intervals indexed by 1. Given any $\lambda_1 \in \bbR_+$, after $\tau_{\rmm,1}^*$ queries, the first estimate $W_1^{(1)}$ is as follows:
\begin{align}
W_1^{(1)} = \max\Big\{j \in [M]:\imath^{h(p)}\big(X_1^{\tau_{\rmm,1}^*}(j);Y_1^{\tau_{\rmm,1}^*}\big) \geq \lambda_1\Big\}.
\end{align}

Through feedback, the oracle learns whether $\tau_{\rmm,1}^*$ is reached at each time during the first estimation. This type of feedback is motivated by stop-feedback strategy in practical communications that does not alter the transmitted symbol beyond telling the transmitter when to stop transmitting. At the time point $\tau_{\rmm,1}^*$, $W_1^{(1)}$ is fed back to the oracle for him to accept or reject it.

\textbf{Hypothesis test phase:}
If $W_1^{(1)}=1$, then the oracle transmits the sequence of $(x_{\rmA}, x_{\rmA}, \ldots)$; otherwise, it sends $(x_{\rmR}, x_{\rmR}, \ldots)$. The questioner constructs the sequential hypothesis test
\begin{align}
H_{\rmA}:Y \sim P_{Y|X=x_{\rmA}}, \\
H_{\rmR}:Y \sim P_{Y|X=x_{\rmR}},
\end{align}
and Wald’s sequential probability ratio test (SPRT)
\begin{align}
\tau_{\text{HT}} := \inf\bigg\{n\geq1:\sum_{i=1}^n\log\frac{P_{Y|X=x_{\rmA}}(Y_i)}{P_{Y|X=x_{\rmR}}(Y_i)}\notin[-a_{\rmR},a_{\rmA}]\bigg\},
\end{align}
where $-a_{\rmR}$ and $a_{\rmA}$ are thresholds of the SPRT. Here, $H_{\rmA}$ and $H_{\rmR}$ correspond to hypothesis to accept and to reject the first estimate $W_1^{(1)}$, respectively.

If $\sum_{i=1}^{\tau_{\text{HT}}}\log\frac{P_{Y|X=x_{\rmA}}(Y_i)}{P_{Y|X=x_{\rmR}}(Y_i)}>a_{\rmA}$, then $H_{\rmA}$ is declared at the time point $\tau_{\rmm,1}^*+\tau_{\text{HT}}$ by the oracle, and the first estimate $W_1^{(1)}$ is accepted. If $\sum_{i=1}^{\tau_{\text{HT}}}\log\frac{P_{Y|X=x_{\rmA}}(Y_i)}{P_{Y|X=x_{\rmR}}(Y_i)}<-a_{\rmR}$, then $H_{\rmR}$ is declared, and the second estimation begins.

\textbf{The second estimation in the non-adaptive query stage:}
The questioner continues to pose the $(\tau_{\rmm,1}^*+1)$-th query, and determines to get the second estimate after a total of $\tau_{\rmm,2}^*$ queries (including the $\tau_{\rmm,1}^*$ queries used for the first estimation). At the time point $\tau_{\rmm,2}^*+\tau_{\text{HT}}$, the questioner get the second estimate $W_2^{(1)}$.

\textbf{The estimation in the cloned adaptive query stage:}
Let $W^{(1)}$ denote the estimate obtained in the non-adaptive query stage, regardless of whether it comes from the first or second estimation. From the questioner’s perspective, the first-level sub-interval containing the target has already been identified as $\calC_{W^{(1)}}^{(1)}$, whereas the eavesdropper still has no information at this stage. In order to quickly localize the target with higher resolution and avoid the leakage of the target location information obtained in the non-adaptive stage, the questioner enters the cloned adaptive query stage. Specifically, the questioner follows the sortPM strategy to generate a query and clones it to all first-level sub-intervals. After $\tau_{\rms}^*$ queries, the questioner can obtain the estimate of second-level sub-interval $W^{(2)}$. Therefore, the questioner knows both the first-level and second-level sub-intervals containing the target, while the eavesdropper can only infer the second-level sub-interval based on the queries he overheard in the cloned adaptive query stage.

\subsubsection{Error Probability}\

Define the following error events:
\begin{align}
\calE_i^{(1)}&:=\big\{W_i^{(1)} \neq 1\big\} ,i\in[2] \\
\calE_{\rmA\rightarrow\rmR}&:=\{\text{The oracle accepts but $H_{\rmR}$ is declared}\} \\
\calE_{\rmR\rightarrow\rmA}&:=\{\text{The oracle rejects but $H_{\rmA}$ is declared}\} \\
\calE_{1 \rightarrow 2}&:=\{\text{The second estimation begins}\} \\
\calE^{(2)}&:=\big\{W^{(2)} \neq 1\big\}.
\end{align}

Then the error probability of the two-stage private query procedure can be bounded as follows:
\begin{align}
&\text{Pr}\Big\{\big(W^{(1)}\neq1\big) \bigcup \big(W^{(2)}\neq1\big)\Big\} \nn\\
&\leq \text{Pr}\big\{W^{(1)}\neq1\big\} + \text{Pr}\big\{W^{(2)}\neq1\big\} \label{err-prob-1}\\
&\leq \text{Pr}\big\{W^{(1)}\neq1\big\} + \varepsilon^{\prime} \label{err-prob-2}\\
&\leq \text{Pr}\Big\{\big(\calE_1^{(1)}\bigcap\calE_{\rmR\rightarrow\rmA}\big) \bigcup \calE_2^{(1)}\Big\} + \varepsilon^{\prime} \label{err-prob-3}\\
&\leq \text{Pr}\big\{\calE_1^{(1)}\big\}\text{Pr}\big\{\calE_{\rmR\rightarrow\rmA}\big\} + \text{Pr}\big\{\calE_2^{(1)}\big\} + \varepsilon^{\prime}, \label{err-prob-4}
\end{align}
where \eqref{err-prob-2} follows from the definition of stopping time for sortPM algorithm in \eqref{tau-s}, and \eqref{err-prob-4} follows from the independence of the events $\calE_1^{(1)}$ and $\calE_{\rmR\rightarrow\rmA}$. Inspired by~\cite[Theorem 1]{zhou2021achievable}, for any $i \in [2]$, the error probabilities of $\text{Pr}\{\calE_i^{(1)}\}$ can be bounded as follows:
\begin{align}
&\text{Pr}\big\{\calE_i^{(1)}\big\} \nn\\
&\leq (L-1)\text{Pr}\Big\{\tau_{\rmm,1}^i(\bX_1^{\infty},Y_1^{\infty},\lambda_i) \geq \tau_{\rmm,2}^i(\bX_1^{\infty},Y_1^{\infty},\lambda_i)\Big\}, \label{err-prob-eventi-1}\\
&\leq (L-1)\exp(-\lambda_i), \label{err-prob-eventi-2}
\end{align}
where $\tau_{\rmm,1}^i(\bX_1^{\infty},Y_1^{\infty},\lambda_i)$ (or $\tau_{\rmm,2}^i(\bX_1^{\infty},Y_1^{\infty},\lambda_i)$) in \eqref{err-prob-eventi-1} denotes the stopping time defined in \eqref{tau-m}, corresponding to the case that the target lies in the first (or second) first-level sub-interval in the $i$-th estimation in the non-adaptive query stage. 

And from~\cite[Theorem 3.1]{woodroofe1982nonlinear}, the error probabilities of $\calE_{\rmR\rightarrow\rmA}$ and $\calE_{\rmA\rightarrow\rmR}$ in the sequential hypothesis test can be bounded as follows:
\begin{align}
&\text{Pr}\big\{\calE_{\rmR\rightarrow\rmA}\big\} \leq \exp(-a_{\rmA}), \label{err-prob-eventRA-1}\\
&\text{Pr}\big\{\calE_{\rmA\rightarrow\rmR}\big\} \leq \exp(-a_{\rmR}). \label{err-prob-eventAR-1}
\end{align}

Therefore, combining \eqref{err-prob-4}, \eqref{err-prob-eventRA-1}, and \eqref{err-prob-eventi-2} we can get the upper bound of the error probability of the two-stage private query procedure as follows:
\begin{align}
&\text{Pr}\Big\{\big(W^{(1)}\neq1\big) \bigcup \big(W^{(2)}\neq1\big)\Big\} \nn\\
&\leq (L-1)\big(\exp(-\lambda_1-a_{\rmA})+\exp(-\lambda_2)\big)+\varepsilon^{\prime}. \label{err-prob-rlt}
\end{align}

\subsubsection{Average Stopping Time}\

Let $N_0$ denote the maximal stopping times of the cloned adaptive query stage, respectively. This procedure consists of two stages of querying, connected by an intermediate stage of sequential hypothesis testing. Hence, the average stopping time of the proposed procedure is given by the sum of the average stopping times across all stages, i.e.,
\begin{align}
&\bbE[\tau] \nn\\
&\leq \bbE[\tau_{\rmm,1}^*] + \text{Pr}\{\calE_{1 \rightarrow 2}\}\bbE[\tau_{\rmm,2}^*-\tau_{\rmm,1}^*] + \bbE[\tau_{\text{HT}}] + \bbE[\tau_{\rms}^*] \label{average-time-1}\\
&\leq \bbE\big[\tau_{\rmm,1}^1(\bX_1^{\infty},Y_1^{\infty},\lambda_1)\big] \nn\\
&+ \text{Pr}\{\calE_{1 \rightarrow 2}\}\bbE\big[\tau_{\rmm,1}^2(\bX_1^{\infty},Y_1^{\infty},\lambda_2)-\tau_{\rmm,1}^1(\bX_1^{\infty},Y_1^{\infty},\lambda_1)\big] \nn\\
&+ \bbE[\tau_{\text{HT}}] + \text{Pr}\big\{W^{(1)}=1\big\}\bbE[\tau_{\rms}^*] + \text{Pr}\{W^{(1)}\neq1\}N_0 \label{average-time-2}\\
&\leq \bbE\big[\tau_{\rmm,1}^1(\bX_1^{\infty},Y_1^{\infty},\lambda_1)\big] \nn\\
&+ \text{Pr}\{\calE_{1 \rightarrow 2}\}\bbE\big[\tau_{\rmm,1}^2(\bX_1^{\infty},Y_1^{\infty},\lambda_2)-\tau_{\rmm,1}^1(\bX_1^{\infty},Y_1^{\infty},\lambda_1)\big] \nn\\
&+ \bbE[\tau_{\text{HT}}] + \bbE[\tau_{\rms,1}(\bX_2^{\infty},Y_2^{\infty},\varepsilon^{\prime})] + \text{Pr}\{W^{(1)}\neq1\}N_0, \label{average-time-3}
\end{align}
where \eqref{average-time-2} follows from the definitions of \eqref{tau-m-star} that $\tau_{\rmm,i}^*\leq\tau_{\rmm,1}^i(\bX_1^{\infty},Y_1^{\infty},\lambda_i)$, and \eqref{average-time-3} follows from the definitions of \eqref{tau-s-star} that $\tau_{\rms,1}^*\leq\tau_{\rms}(\bX_2^{\infty},Y_2^{\infty},\varepsilon^{\prime})$.

Then we bound the probability that the second estimation begins as follows:
\begin{align}
\text{Pr}\{\calE_{1 \rightarrow 2}\} &= \text{Pr}\Big\{ \Big(\calE_1^{(1)}\bigcap\calE_{\rmR\rightarrow\rmA}^{\rmc}\Big)\bigcup\Big((\calE_1^{(1)})^{\rmc}\bigcap\calE_{\rmA\rightarrow\rmR}\Big) \Big\} \\
&\leq \text{Pr}\big\{\calE_1^{(1)}\big\} + \text{Pr}\big\{\calE_{\rmA\rightarrow\rmR}\big\} \\
&\leq (L-1)\exp(-\lambda_1) + \exp(-a_{\rmR}), \label{err-prob-1to2}
\end{align}
where $\calE^{\rmc}$ denotes the complement of event $\calE$.

Recall the definitions of $b(\cdot)$ in \eqref{b-func} and $D(\cdot)$ in \eqref{KL-div} in the Notation part, and $C$ in \eqref{capacity}, respectively. From~\cite[Lemma 1]{yavas2025variable}, we can bound the rest terms in \eqref{average-time-3} as follows:
\begin{align}
&\bbE[\tau_{\text{HT}}|H_{\rmA}] \leq \frac{a_{\rmA}+b_{\rmA}}{D(P_{Y|X=x_{\rmA}}\|P_{Y|X=x_{\rmR}})}, \label{expectation1}\\
&\bbE[\tau_{\text{HT}}|H_{\rmR}] \leq \frac{a_{\rmR}+b_{\rmR}}{D(P_{Y|X=x_{\rmR}}\|P_{Y|X=x_{\rmA}})}, \label{expectation2}\\
&\bbE[\tau_{\text{HT}}] \nn\\
&= \text{Pr}\{H_{\rmA}\}\bbE[\tau_{\text{HT}}|H_{\rmA}]+\text{Pr}\{H_{\rmR}\}\bbE[\tau_{\text{HT}}|H_{\rmR}] \\
&\leq \bbE[\tau_{\text{HT}}|H_{\rmA}]+\text{Pr}\{H_{\rmR}\}\bbE[\tau_{\text{HT}}|H_{\rmR}] \\
&\leq \bbE[\tau_{\text{HT}}|H_{\rmA}]+(L-1)\exp(-\lambda_1)\bbE[\tau_{\text{HT}}|H_{\rmR}], \label{expectation3}\\
&\bbE[\tau_{\rmm,1}^1(\bX_1^{\infty},Y_1^{\infty},\lambda_1)] \leq \frac{\lambda_1+b}{C}, \label{expectation4}\\
&\bbE\big[\tau_{\rmm,1}^2(\bX_1^{\infty},Y_1^{\infty},\lambda_2)-\tau_{\rmm,1}^1(\bX_1^{\infty},Y_1^{\infty},\lambda_1)\big] \leq \frac{\lambda_2-\lambda_1+b}{C}, \label{expectation5}
\end{align}
where $b_{\rmA}$, $b_{\rmR}$ and $b$ are set to $b_{\rmA}(P_{\log\frac{P_{Y|X=x_{\rmA}(Y_{\rmA})}}{P_{Y|X=x_{\rmR}(Y_{\rmA})}}})$, $b_{\rmR}=b(P_{\log\frac{P_{Y|X=x_{\rmR}(Y_{\rmR})}}{P_{Y|X=x_{\rmA}(Y_{\rmR})}}})$ and $b=b(P_{\imath^{h(p)}(X;Y)})$, respectively, and \eqref{expectation3} follows from replacing $\text{Pr}\{H_{\rmR}\}$ with $\text{Pr}\{\calE_1^{(1)}\}$.

Therefore, combining \eqref{err-prob-4}, \eqref{average-time-3}, \eqref{err-prob-1to2} and \eqref{expectation1}-\eqref{expectation5}, we have the upper bound of the average stopping time of the two-stage private query procedure as follows:
\begin{align}
&\bbE[\tau] \nn\\
&\leq \frac{\lambda_1+b}{C}+\big((L-1)\exp(-\lambda_1) + \exp(-a_{\rmR})\big)\frac{\lambda_2-\lambda_1+b}{C} \nn\\
&+ \frac{a_{\rmA}+b_{\rmA}}{D(P_{Y|X=x_{\rmA}}\|P_{Y|X=x_{\rmR}})} \nn\\
&+ (L-1)\exp(-\lambda_1)\frac{a_{\rmR}+b_{\rmR}}{D(P_{Y|X=x_{\rmR}}\|P_{Y|X=x_{\rmA}})} \nn\\
&+ \bbE\big[\tau_{\rms,1}(\bX_2^{\infty},Y_2^{\infty},\varepsilon^{\prime})\big] \nn\\
&+ (L-1)\big(\exp(-\lambda_1+a_{\rmA})+\exp(-\lambda_2)\big)N_0. \label{average-time-rlt}
\end{align}

From the right-hand sides of \eqref{err-prob-rlt} and \eqref{average-time-rlt}, let
\begin{align}
&\bar{\varepsilon} = (L-1)\big(\exp(-\lambda_1-a_{\rmA})+\exp(-\lambda_2)\big)+\varepsilon^{\prime}, \label{bar-var}\\
&N_1 \nn\\
&= \frac{\lambda_1+b}{C}+\big((L-1)\exp(-\lambda_1) + \exp(-a_{\rmR})\big)\frac{\lambda_2-\lambda_1+b}{C} \nn\\
&+ \frac{a_{\rmA}+b_{\rmA}}{D(P_{Y|X=x_{\rmA}}\|P_{Y|X=x_{\rmR}})} \nn\\
&+ (L-1)\exp(-\lambda_1)\frac{a_{\rmR}+b_{\rmR}}{D(P_{Y|X=x_{\rmR}}\|P_{Y|X=x_{\rmA}})}, \label{N1}\\
&N_2 = \bbE\big[\tau_{\rms,1}(\bX_2^{\infty},Y_2^{\infty},\varepsilon^{\prime})\big] \nn\\
&+ (L-1)\big(\exp(-\lambda_1+a_{\rmA})+\exp(-\lambda_2)\big)N_0, \label{N2}\\
&\bar{N} = N_1+N_2. \label{barN-Nd}
\end{align}
From the above analysis, there exists an $(\bar{N},L,1/M,\bar{\varepsilon})$ query procedure. We use the stop-at-time-zero strategy described in~\cite{polyanskiy2011feedback}. Let $N$ and $\varepsilon$ denote the average number of queries and the average error probability of the described private query procedure obtained by this time-sharing strategy, i.e.,
\begin{align}
N &\leq (1-\varepsilon_0)\bar{N}, \label{N-barN}\\
\varepsilon &\leq \varepsilon_0+(1-\varepsilon_0)\bar{\varepsilon}. \label{eps-bareps}
\end{align}

\subsection{Proof of the Asymptotic Bound (Theorem \ref{2order-asymptotic})} \label{appendix_b}

Given $L \in \bbN$ and $L \geq 2$, let $N_{\dagger}$ is the solution to the following equality
\begin{align}
\log L=CN_{\dagger}-\log\log N_{\dagger}. \label{choice-Ndagger}
\end{align}
which denotes an upper bound on the number of queries of the first estimation in the non-adaptive query stage. To derive the asymptotic bound, we choose
\begin{align}
\lambda_1 &= \log L + \log\log N_{\dagger}, \label{choice-lambda1}\\
\lambda_2 &= \log L + \log N_{\dagger}, \label{choice-lambda2}\\
a_{\rmA} &= a_{\rmR} = \log N_{\dagger}. \label{choice-a}
\end{align}
Recall the definition of $\widetilde{C}$ in \eqref{capacity-wt}. When $N_{\dagger} \rightarrow \infty$, the number of queries $N_1$ in \eqref{N1} in the non-adaptive query procedure is as follows:
\begin{align}
&N_1 \nn\\
&\leq \frac{\lambda_1+b}{C}+\big((L-1)\exp(-\lambda_1) + \exp(-a_{\rmR})\big)\frac{\lambda_2-\lambda_1+b}{C} \nn\\
&+ \frac{a_{\rmA}+b_{\rmA}}{D(P_{Y|X=x_{\rmA}}\|P_{Y|X=x_{\rmR}})} \nn\\
&+ (L-1)\exp(-\lambda_1)\frac{a_{\rmR}+b_{\rmR}}{D(P_{Y|X=x_{\rmR}}\|P_{Y|X=x_{\rmA}})} \\
&= N_{\dagger}+\frac{\log N_{\dagger}}{\widetilde{C}}+O(1). \label{n-dagger}
\end{align}

Since the questioner knows the estimate of the non-adaptive query stage, the cloning operation does not have a significant impact on the asymptotic analysis. The number of queries $N_2$ in \eqref{N2} in the cloned adaptive query procedure is similar to~\cite[Theorem 1]{chiu2021low} and~\cite[Remark 4]{naghshvar2015extrinsic} as follows:
\begin{align}
N_2 &\leq \bbE\big[\tau_{\rms,1}(\bX_2^{\infty},Y_2^{\infty},\varepsilon^{\prime})\big] \nn\\
&+ (L-1)\big(\exp(-\lambda_1+a_{\rmA})+\exp(-\lambda_2)\big)n_2 \\
&\leq \bbE\big[\tau_{\rms,1}(\bX_2^{\infty},Y_2^{\infty},\varepsilon^{\prime})\big] + O(1) \\
&\leq \frac{\log(M/L)}{C}+\frac{\log(1/\varepsilon^{\prime})}{\widetilde{C}}+o\Big(\log\frac{M}{\varepsilon^{\prime}L}\Big).
\end{align}

By choosing $M$ such that
\begin{align}
\log\frac{M}{L}=\log M-\log L=CN_2-\frac{C}{\widetilde{C}}\log\frac{1}{\varepsilon^{\prime}}, \label{logML}
\end{align}
we have that
\begin{align}
\log M = CN_{\dagger}+CN_2-\log\log N_{\dagger}-\frac{C}{\widetilde{C}}\log\frac{1}{\varepsilon^{\prime}}+O(1). \label{logM}
\end{align}

From \eqref{n-dagger}, we get
\begin{align}
N_{\dagger} &= N_1-\frac{\log N_{\dagger}}{\widetilde{C}}-O(1), \label{Nd-to-N1-1}\\
\log N_{\dagger} &= \log\Big(N_1-\frac{\log N_{\dagger}}{\widetilde{C}}-O(1)\Big) \\
&= \log N_1 + \log\Big(1-\frac{\log N_{\dagger}}{\widetilde{C}N_1}-O(1)\Big) \\
&= \log N_1 + O(1). \label{Nd-to-N1-2}
\end{align}
Combining \eqref{logM}, \eqref{Nd-to-N1-1} and \eqref{Nd-to-N1-2}, we have that
\begin{align}
\log M &= CN_1+CN_2-\frac{C}{\widetilde{C}}\log N_1 \nn\\
&-\log\log N_1-\frac{C}{\widetilde{C}}\log\frac{1}{\varepsilon^{\prime}}+O(1) \\
&= C\bar{N} -\frac{C}{\widetilde{C}}\log N_1 -\log\log N_1-\frac{C}{\widetilde{C}}\log\frac{1}{\varepsilon^{\prime}}+O(1). \label{logM-rlt-1}
\end{align}

Combining \eqref{bar-var}, \eqref{choice-Ndagger}, \eqref{choice-lambda1}, \eqref{choice-lambda2} and \eqref{choice-a}, we have that
\begin{align}
\bar{\varepsilon} &\leq (L-1)\big(\exp(-\lambda_1-a_{\rmA})+\exp(-\lambda_2)\big)+\varepsilon^{\prime} \\
&\leq \frac{1}{N_{\dagger}}\Big(1+\frac{1}{\log N_{\dagger}}\Big)+\varepsilon^{\prime}.
\end{align}
Let $B(N_{\dagger}):=\frac{1}{N_{\dagger}}(1+\frac{1}{\log N_{\dagger}})$. From \eqref{eps-bareps}, we have that
\begin{align}
\varepsilon_0 \geq \frac{\varepsilon-\varepsilon^{\prime}-B(N_{\dagger})}{1-\varepsilon^{\prime}-B(N_{\dagger})}.
\end{align}
So we choose
\begin{align}
\varepsilon_0 = \frac{\varepsilon-B(N_{\dagger})}{1-B(N_{\dagger})} = \varepsilon-\Omega(B(N_{\dagger})). \label{varepsilon0}
\end{align}

Finally, from \eqref{N-barN} and \eqref{varepsilon0}, we have that
\begin{align}
\bar{N} &= \frac{1}{1-\varepsilon}N. \label{barN-to-N}
\end{align}
Therefore, combining \eqref{logM-rlt-1} and \eqref{barN-to-N}, we can obtain the asymptotic bound as follows:
\begin{align}
&\log M \nn\\
&= \frac{1}{1-\varepsilon}CN-\frac{C}{\widetilde{C}}\log N_1-\log\log N_1-\frac{C}{\widetilde{C}}\log\frac{1}{\varepsilon^{\prime}}+O(1),
\end{align}
where $N_1$ is the solution to the following equality
\begin{align}
\log L = CN_1-\frac{C}{\widetilde{C}}\log N_1-\log\log N_1.
\end{align}

\end{document}